\documentclass[aps,showpacs,twocolumn,amsmath,preprintnumbers,nofootinbib,floatfix,secnumarabic]{revtex4}

\usepackage{epsfig,verbatim}
\usepackage{amssymb,amsfonts}
\usepackage{graphicx}

\usepackage[caption=false]{subfig}
\def\sct{0.51} 

\usepackage{color}

\def\jrn#1#2#3#4#5#6{#3 \textbf{#4}, #5 (#6)}  

\def\scn#1#2{\section{#1}\lb{#2}}
\def\sscn#1#2{\subsection{#1}\lb{#2}}
\def\bfl{\begin{flushleft}}
\def\efl{\end{flushleft}}
\def\bfr{\begin{flushright}}
\def\efr{\end{flushright}}
\def\bc{\begin{center}}
\def\ec{\end{center}}
\def\be{\begin{equation}}
\def\ee{\end{equation}}
\def\bse{\begin{subequations}}
\def\ese{\end{subequations}}
\def\ba{\begin{eqnarray}}
\def\ea{\end{eqnarray}}
\def\baa#1{\begin{array}{#1}}
\def\eaa{\end{array}}
\def\bw{\begin{widetext}}
\def\ew{\end{widetext}}
\def\nn{\nonumber }
\def\lb#1{\label{#1}}
\def\bit{\begin{itemize}}
\def\eit{\end{itemize}}
\def\bco{}
\def\bcs{\begin{cases}}
\def\ecs{\end{cases}}

\def\lapl{\Delta_D}

\def\en{{E}}

\def\becc{\lambda}
\def\nbod{{\cal N}}

\begin{document}

\preprint{\small Z. Naturforsch. A  72 (2017) 677-687
\qquad DOI: 10.1515/zna-2017-0134
\ [\url{http://doi.org/10.1515/zna-2017-0134}]
}

\title{
Stability and metastability of trapless Bose-Einstein condensates and quantum liquids
}

\author{Konstantin G. Zloshchastiev}
\affiliation{Institute of Systems Science, Durban University of Technology, P.O. Box 1334, Durban 4000, South Africa}


\begin{abstract} 
Various kinds of Bose-Einstein condensates are considered,
which evolve 
without any geometric constraints or external trap potentials including gravitational.
For studies of their collective oscillations and stability,
including the metastability and macroscopic tunneling phenomena, 
both the variational approach and the Vakhitov-Kolokolov criterion are employed; 
calculations are done for condensates of an arbitrary spatial dimension.
It is determined that that the trapless condensate described by the logarithmic wave equation is essentially stable, regardless
of its dimensionality,
while the trapless condensates described by wave equations of a polynomial type with respect to the wavefunction, 
such as the Gross-Pitaevskii (cubic), cubic-quintic, and so on, are at best metastable.
This means that trapless ``polynomial'' condensates are unstable against spontaneous delocalization caused by
fluctuations of their width, density and energy, leading to a finite lifetime. 
\end{abstract}

\date{8 Dec 2016 [APS], 18 Apr 2017 [ZNA]}


\maketitle

\section{Introduction}

Studies of collective excitations inside  Bose-Einstein condensates (BEC) 
and their stability are an important direction of research, which has direct connections
to both experimental studies and technological applications.
Historically, most of the research done was primarily focused on cold gases in which
two-body interactions are predominant; therefore they can be described by
the Gross-Pitaevskii equation (GPE) which is cubic with respect to a
condensate's wavefunction and thus also known as the cubic Schr\"odinger equation.
Naturally, confining those gases requires some kind of external (trapping) potential,
such as a harmonic one.
Stability studies of those systems are based on a formalism developed 
by Zakharov and collaborators \cite{zak72,zak72c,wei83,ber98},
extended for cases where a trap is included \cite{r23,sto97,r6,r7,bak00,r8,r9}.
It was demonstrated that trap potentials facilitate stability of the GPE BEC by suppressing
the collapse process, 
which is still inevitable
if
the number of atoms exceeds a certain critical limit.
Moreover, when studying the collapse processes in diluted condensates, one should bear in mind that,
once the density of a condensate rises above a certain threshold, the two-body approximation
can become too crude, as will be discussed below in more detail.

Notwithstanding the phenomenological success of the Gross-Pitaevskii approximation for some systems, 
it soon became apparent that there are condensates for which three-body interactions play an important role 
(e.g., when density rises or when the two-body
interaction gets switched off by tuning external fields)  \cite{r10,r11,r12,r13,r14,r15,r16}. 
In particular, adding a three-body interaction can considerably increase the BEC's stability region \cite{r14}. 
Another example of where multi-body (three and more) interactions become
very important for forming bound states of bosons at low temperatures is the Efimov
state, which has been experimentally observed \cite{efim,efimb,efimc,efimexp,nfj98}.
It seems also that multi-body interactions are inevitable for explaining recent experimental
data \cite{khm17}, which are currently receiving a disputable interpretation.
Moreover, the mere notion of a general Bose-Einstein condensate itself presumes,
strictly speaking, that correlations are 
simultaneously established between all the particles which form a condensate, not just between a
pair of them.
It is thus natural 
to go beyond the two-, three-, or even few-body, interactions' approximations, and
consider wave equations' terms containing \textit{all} powers of a condensate wavefunction.
In turn, this leads to an appearance of transcendental functions in wave equations.

In the works \cite{Zloshchastiev:2009zw,r17,r20,r21}, a new quantum Bose liquid was proposed,
which is described by a nonlinear quantum wave equation of a logarithmic kind,
previously introduced on different grounds by Rosen and Bialynicki-Birula and Mycielski \cite{r18,r18a,r19,r19b,bbm79}.
Currently, applications of this equation, both in its Euclidean and Lorentz-symmetric versions, can be found in
nonlinear scalar field theory \cite{r18,r18a},
extensions of quantum mechanics \cite{r19},
physics of particles in presence of nontrivial vacuum
\cite{r17,gg14,Dzhunushaliev:2012zb,gul14,gul15,dmz15}, 
microscopical theory of superfluidity of helium II \cite{r21},
optics and transport or diffusion phenomena \cite{gg2,gg5,gg6}, 
nuclear physics \cite{gg3,gg4}, 
and
theory of dissipative systems and quantum information \cite{gg7,gg8,gg9,gg10,gg11,gg12}. 
Moreover, applications of logarithmic wave equations
can be also found in classical and quantum gravity \cite{Zloshchastiev:2009zw,r17,szm16},
where one can utilize the fluid/gravity correspondence between nonrelativistic
inviscid fluids (such as superfluids) and pseudo-Riemannian manifolds \cite{unr81,unr81a,vis98,zlo99,vol03}.

While some of the above results are still a subject for future experimental verification, 
one practical application of a logarithmic
fluid model is immediately apparent.
In Ref. \cite{r20}, the fluid
was shown to be a proper superfluid, \textit{i.e.}, 
a quantum Bose liquid which simultaneously possesses the following two properties: 
its spectrum of excitations 
has a Landau's ``roton'' form, which 
guarantees that dissipation is suppressed at microscopic level, and 
its macroscopic (averaged) equation of state has, in the leading approximation at least,
an ideal-fluid form (a ratio of density to pressure 
is constant),
which guarantees the perfectly elastic collisions.
More specifically, in Ref. \cite{r21}, the fluid was used to construct a fully analytical theory 
of a superfluid component of helium II,
which has a very good agreement with an experiment:
with only one essential (non-scale) parameter to fit, 
this model could theoretically explain the main three properties
of helium II --  ``roton'' spectrum, structure factor, and speed of sound --
with high accuracy.

Altogether, in Refs.~\cite{r20,r21,bo15}, it has been shown that the logarithmic condensate
has features that are drastically different from those of the Gross-Pitaevskii condensate.
In this paper, we continue this line of research:
we study the stability
properties of the logarithmic BEC in free space
related to its collective excitations  (in Ref. \cite{bo15} these were considered for the BEC in a harmonic trap). 

Our paper is organized as follows. 
In Sec. \ref{s:nst}, we discuss different notions of stability and set definitions for subsequent sections.
In Sec. \ref{s:ini}, we present the basics of a theory of logarithmic BEC in absence of any trap potentials and geometrical constraints.
In Sec. \ref{s:ss}, an analysis of collective oscillations and stability  is performed, using both the variational approach and the Vakhitov-Kolokolov (VK) criterion.
In Sec. \ref{s:poly}, a comparison between collective oscillations and stability 
for the logarithmic BEC and ``polynomial'' condensates with few-body interactions
(including the two- and three-body ones) is done.
Concluding remarks and discussions are provided in Sec. \ref{s:con}.

\scn{Stability analysis of quantum liquids and gases}{s:nst}

As a starting point, we emphasize that
our stability studies should not be confused with a conventional stability analysis of optical solitons 
and other objects whose evolution is also governed by nonlinear wave equations \cite{ab79,abl79,and83,alr88,fe96,fe96a}.
The underlying physics of quantum molecular gases and liquids is very different from that of optical fibers and other electromagnetic (EM) materials.
For instance, for optical solitons, the wave equation solution  is originally a deterministic function 
related to a strength of EM wave field and governed by Maxwell equations
(although some dissipative effects for EM waves in media can be described using quantum statistics \cite{z16prb,z17adp}),
whereas for quantum condensates it is \textit{a priori} a proper wavefunction which has a quantum-mechanical probabilistic interpretation.
The latter leads to the occurrence of many kinds of quantum effects in condensates, including the macroscopic tunneling phenomenon, which will be further discussed below.
These effects obviously affect the stability properties of quantum liquids and gases, a factor which often causes stability methods and results to differ from those for other classes of physical objects.

In a theory of quantum liquids and gases including Bose-Einstein condensates, there are a few kinds of stability, each with its own definitions and criteria. 
The reason for this variety lies in the complex nature of Bose-Einstein condensation: condensates are not only nonlinear phenomena described by solutions of nonlinear wave equations (e.g. the  Gross-Pitaevskii one); 
but also essentially quantum systems whose states can be eigenstates of wave equations, superpositions of eigenstates, or even statistical ensembles of eigenstates (\textit{i.e.}, mixed states).
Their quantum nature implies that condensates not only follow
continuous evolution of eigenstates governed by wave equations
but can also experience spontaneous transitions between these eigenstates,
due to the inevitable presence of quantum fluctuations.

The most common types of stability analysis are the linear and orbital stability
of a given solution, such
as the Vakhitov-Kolokolov criterion \cite{vk73}, which formalism comes from a classical theory of stability of nonlinear systems. According to this approach, one takes a fixed solution of a wave equation and studies its linearized response to small perturbations.
From the viewpoint of a quantum theory, this solution describes a state with a definite value of energy which is kept fixed during a perturbation. Therefore, the VK-type stability analysis is essentially classic (although some modifications are probably possible): for instance, it does not take into account effects from possible quantum transitions between levels. 
In the quantum realm, these transitions can happen, \textit{e.g.}, if a state corresponding to a given solution is not a ground one: for instance, a solution which is stable from a classical point of view, may be unstable against spontaneous transitions if it corresponds to an excited state.

The other stability approach, variational, which was initially proposed for studies of optical solitons \cite{and83,alr88},
does not initiate from the fixed solution of a system. 
Instead, similar to a Ritz optimization procedure, one starts with trial functions and searches for a configuration which minimizes a field-theoretical action of the condensate (related to an average energy),
then one studies linear oscillations near such minimum. 
If such oscillations do not blow up with time, then a system stays near the above-mentioned minimum, therefore it is stable. By its construction, this method does not \textit{ab initio} fix an energy level, therefore it can detect instability against quantum transitions between different states which can occur spontaneously.

In what follows, we will be using these two notions of stability, where possible: by stability and metastability analysis of condensates we will understand the variational method taking into account quantum effects, while the linear and orbital stability approach of a given state will be implemented using the Vakhitov-Kolokolov criterion.

\scn{Trapless logarithmic BEC: Basics}{s:ini}

Let us consider a system of quantum Bose
particles, with a fixed mean number  $N$, which evolves in a $D$-dimensional Euclidean space,
in the absence of any external potentials
including gravity. 
We assume that the kinetic energy of the particles is sufficiently low
for the Bose-Einstein condensate to form.
If interparticle interactions are sufficiently large (which can happen, e.g., 
at large densities), 
the dynamics of such BEC
can be described by the logarithmic Schr\"odinger equation (LogSE):
\begin{equation}\label{e1}
i\hbar \partial_t \Psi
=
\left[-\frac{\hbar^{2}}{2 m} \lapl
-b
 \ln(a^{D} |\Psi|^{2})
\right]\Psi
.
\end{equation}
where 
$\Psi=\Psi(\vec r, t)$ is the condensate wavefunction normalized to the particle number $N$,
\be\lb{norm}
\left\langle \Psi | \Psi\right\rangle
\equiv
\int |\Psi|^2 d^D \vec r  = N,
\ee
and
$\lapl = \vec{\nabla}_D \cdot \vec{\nabla}_D$ is the $D$-dimensional Laplacian. 
The parameter $b$ measures the strength of a nonlinear interaction,
$a$ is a parameter of dimensionality length required to make the argument of the logarithm dimensionless,
and $m$ is a particle's mass.

\begin{figure}[htbt]
\begin{center}\epsfig{figure=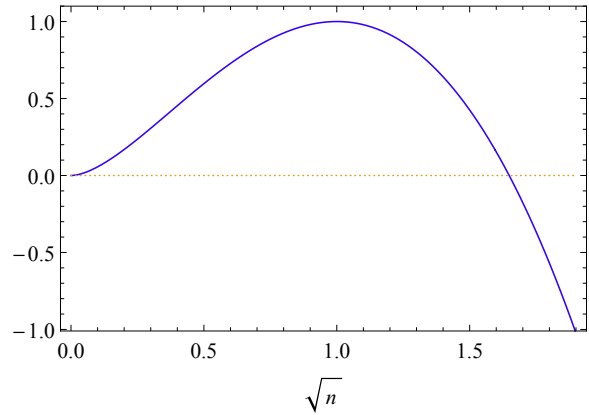,width=  1.0\columnwidth}\end{center}
\caption{
Field-theoretical potential density (\ref{e:ftpot}) of the logarithmic BEC (in units $b/{a}^{D}$) versus 
$| \Psi |$
(in units ${a}^{-D/2}$).
Due to the normalization constraint (\ref{norm}), this plot must be viewed as effectively having been
placed 
inside a well with an infinite wall starting at $n \geqslant n_\text{cut}$,  $n_\text{cut}$ 
being some finite number.  
}
\label{f:ftpot}
\end{figure}

Equation (\ref{e1}) can also be derived, as an Euler-Lagrange equation,
from a field-theoretical
action, where the Lagrangian density is
given by
\be\lb{ftlagrlog}
{\cal L}
 =
\frac{i\hbar}{2}(\Psi \partial_t\Psi^* - \Psi^*\partial_t\Psi)+
\frac{\hbar^2}{2 m}
|\vec\nabla \Psi|^2
+
V (|\Psi|^2)
,
\ee
where the field-theoretical potential density (not to be confused with the
external trap potential) is defined as
\be\lb{e:ftpot}
V (n) = -
b\,
n
\left[
\ln{(n  a^D)} -1
\right]
,
\ee
for $n = |\Psi|^2$.
For positive values of $b$,
this potential opens down and has local non-zero maxima at $n_\text{ext} =  1/a^{D}$, see Fig. \ref{f:ftpot}.
In spite of the fact that it is not bounded from below as a function 
of $\sqrt n = |\Psi|$, no particle density divergences arise since the condensate wavefunction cannot take
arbitrarily large values, due to the constraint (\ref{norm}), as discussed in Ref. \cite{r21}.

From Fig. \ref{f:ftpot}, one can easily see that the field-theoretical potential of a logarithmic BEC
changes its sign when its particle density crosses a certain value, $n_0 = \text e/a^D = \text e \, n_\text{ext} $.
This switching between attraction and repulsion depending on a size can be used for explaining recent experimental data \cite{khm17}, which indicate the presence of some localization mechanism even for low-density condensates.
Indeed, below we will demonstrate that this feature manifests itself in 
the condensate's
stability against both collapse
and unbounded expansion.

In principle,
one could perform a Taylor series expansion of the non-linear part $\ln(a^{D} |\Psi|^{2})$
around some point,
and obtain in the lowest order
the Gross-Pitaevskii equation, in the
next-to-lowest order -- the cubic-quintic Schr\"odinger equation (CQSE), 
and in higher orders -- the higher-degree polynomial terms.
All these terms
describe interactions of a finite amount of particles at any given instant of time.
Therefore, one might assume that the properties of the logarithmic model would be
similar to those models, at least qualitatively, and the polynomial models' properties would 
be able to reproduce all features of the logarithmic model by considering sufficiently
many terms in the series expansion.
However, this turns out to be incorrect: by restricting to any \textit{finite} number of terms in series, one drastically alters the main properties of a corresponding condensate model, as we will see in the next sections.
Therefore, a nonperturbative treatment is essential when dealing with 
the ``transcendental'' condensates in general and the logarithmic ones in particular.

Furthermore, the system (\ref{e1})
has two natural scales of length,
two of time, and two of mass:
\be
\text{L} 
= \left\{|a|,  \frac{\hbar}{\sqrt{m |b|}} \right\}
,
\
\text{T} 
= \left\{\frac{\hbar}{|b|}, \frac{m a^2}{\hbar} \right\}
,\
\text{M} 
= \left\{m,  \frac{\hbar^2}{a^2 |b|} \right\}
,
\ee
which can be used to obtain dimensionless quantities.
Assuming $a, b>0$ and
\be
\vec r \,'= \vec r/\ell,\
t' = t/\tau
,\ 
\psi = a^{D/2} \Psi
,
\ee
where
\be
\ell = \hbar/\sqrt{m b}, \
\tau = \hbar / b
,
\ee
we can write Eq. (\ref{e1}) in a dimensionless form.
From Eqs. (\ref{e1}) and (\ref{norm})
we obtain, respectively:
\begin{equation}\label{e5}
i \partial_{t'} \psi
+
\frac{1}{2}\lapl^\prime \psi
+
 \ln(|\psi|^{2}) \psi
=
0
,
\end{equation}
and
\begin{equation}\label{e4}
\left\langle \psi | \psi\right\rangle'
\equiv
\int |\psi|^2 d^D \vec{r}\,'  
=
N (a/\ell)^{D} \equiv \bar N
,
\end{equation}
where $\bar N$ will be called the reduced number of particles.

Furthermore, the action that generates Eq.~(\ref{e5}) 
can be written in a dimensionless form as
\be\lb{action}
S [\psi,\psi^*] = 
\int \mathcal{L}' d t' d^D \vec{r}^\prime
=
\int L' d t'
,
\ee
where the
dimensionless Lagrangian density is given by
\begin{equation}\label{e10gen}
\mathcal{L}'
=\frac{i}{2}(\psi \partial_{t'}\psi^*-\psi ^*\partial_{t'}\psi)+\frac{1}{2}|\vec{\nabla}_D^\prime \psi|^{2}
-|\psi|^{2}(\ln(|\psi|^{2})-1)\,,\end{equation}
and
the dimensionless chemical potential $\mu' = \mu/b$ is given as an eigenvalue 
of a stationary version of Eq.~(\ref{e5}),
\begin{equation}\label{e7}
\lapl^\prime \phi
+ 
\left(
\mu'
+
4 \ln{|\phi|}
\right)\phi
=0
,
\end{equation}
where $\psi(\vec{r}\,', t') = \exp[-i(\mu' t'/2)]\phi(\vec{r}\,')$.


\scn{Trapless logarithmic BEC: Stability}{s:ss}

Let us consider a spherically symmetric configuration of freely moving logarithmic BEC in a $D$-dimensional Euclidean space. 
For example, in a case $D=3$ 
this symmetry is the most natural one that can arise in absence of any trapping potentials including gravity.
Note that, although a logarithmic BEC in a harmonic trap
was considered in Ref. \cite{bo15}, 
those results cannot be directly applied to a trapless case -- because
in a zero trap frequency limit, the parametrization used becomes singular.
Thus, for a trapless BEC, we have to start our calculations anew.

In this section, we will be omitting primes, assuming instead that length is measured in
units $\ell$, time - in units $\tau$, energy - in units $b$, and so on. 
For the stability analysis of the logarithmic condensate we will employ the following two approaches.

\sscn{Variational approach}{s:ssva}

In order to analyze the dynamics of logarithmic condensate, it is
convenient to follow a variational approach \cite{r23,sto97,r7,r24,r22,r28}. 
We will seek the solutions of Eq.~(\ref{e5}) using the trial functions 
\begin{equation}\label{e8}
\psi(r,t)=A \exp[-\frac{r^2}{2\xi^{2}}+i \beta r^2 +i \alpha]\,,
\end{equation}
where $A=A(t)$ is the amplitude, $\xi = \xi(t)$ is the width, $\alpha = \alpha(t)$ is the
linear phase of the condensate, and $\beta = \beta(t)$ is the chirp parameter \cite{r24}; 
these functions become \textit{de facto} the collective degrees of freedom of the condensate.
The integral over the whole space can be transformed into
\ba
\int  d^D \vec{r}
&=&
\left\{
\baa{lll}
\frac{2 \pi^{D/2}}{\Gamma(D/2)}
\int\limits_0^{\infty }  d r r^{D-1} &\text{if}&D>1,\\
\int\limits_{-\infty}^{\infty}  d r&\text{if}&D=1,
\eaa
\right.
\lb{intdef}
\ea
where 
$\Gamma(x)$ is the Euler Gamma function.
Therefore, the (reduced) number of particles (\ref{e4}) can be computed as
\begin{equation}\label{e9}
\bar N =\pi^{D/2} A^2 \xi^D
= \text{const}
,
\end{equation}
whereas the averaged Lagrangian can be derived, using 
Eqs. (\ref{action}), (\ref{e10gen}), (\ref{e8}) and (\ref{intdef}), 
as
\be
L
=
\pi^{D/2}A^2\xi^{D}
\Big[
1
+
\frac{D}{2}
+
\frac{D}{4\xi^2}
+\frac{D}{2} \xi^2\left(\dot{\beta}+2\beta^2\right)
+\dot{\alpha}
- \ln A^2
\Big]
,
\label{e11}
\ee
where 
dot represents a time derivative.
By analyzing the corresponding Euler-Lagrange equations,
$
\frac{\partial L}{\partial q}-\frac{d}{d t}\frac{\partial L}{\partial\dot{q}}=0
$,
where $q = \{A(t), \xi(t),\alpha(t),\beta(t) \}$, 
we obtain, after some rearrangement,
\ba
&&
\dot{\alpha}
+\frac{D+2}{2}
\left[
1
+
\xi^2 \left(\dot{\beta} + 2 \beta ^2\right)
\right]
+\frac{D-2}{4 \xi ^2}
=
\ln{A^2}\!,~~~
\label{e14}\\&&
\dot{\alpha}
+\frac{D}{2}
\left[
1
+
\xi^2 \left(\dot{\beta} + 2 \beta ^2\right)
+
\frac{1}{2 \xi ^2}
\right]
=
\ln{A^2}
,\label{e15}\\&&
\dot{\xi}
-
2\xi\beta
=
0
, \label{e16}
\ea
together with Eq. (\ref{e9}).
Furthermore, these equations can be rewritten in the form
\ba
A^2 &=&
\frac{\bar N}{\pi^{D/2} \xi^D}
,\lb{e18a}\\
\alpha
&=&
\alpha_0
+
\ln{\!\left( \frac{\bar N}{\pi^{D/2}} \right)}\,
t
-
\frac{D}{2}
\int
\left(
\ln{\xi^2}
+
\frac{1}{\xi ^2}
\right)
d t
,~~~\\
\beta
&=&
\frac{\dot{\xi}}{2 \xi}
,
\ea
and
\be
\ddot{\xi}
+\frac{2}{\xi}
-\frac{1}{\xi^{3}}
=0
, \label{e18}
\ee
where $\alpha_0$ is an integration constant.
The equations reveal that the evolution equation for width $\xi$ is a core equation 
of the system's dynamics, and also that the amplitude and linear phase of the condensate are 
generally $D$-dependent, whereas the width and chirp do not depend on the dimensionality of
the condensate.

Furthermore, for the wavefunction (\ref{e8}), using Eqs. (\ref{e18a})-(\ref{e18}),  one can derive that 
\ba
\left\langle r^2 \right\rangle 
&=& 
\frac{1}{2} D 
\xi^2
,\\
\left\langle p^2 \right\rangle 
&=& 
\frac{1}{2} \hbar^2 D 
\left(
\dot\xi^2
+
\frac{1}{\xi^2}
\right)
,\\
\left\langle E \right\rangle 
&=& 
\frac{1}{2} \hbar D 
\left[
\frac{1}{2} \dot\xi^2
+
\frac{1}{2 \xi^2}
+
\ln{\left(
\frac{\pi \xi^2}{\bar N^{2/D}}
\right)}
+ 1
\right]
\nn\\
&=&
\frac{1}{2 \hbar}
\left\langle p^2 \right\rangle 
+
\frac{1}{2} \hbar D 
\left[
\ln{\left(
\frac{2 \pi  \left\langle r^2 \right\rangle }{D \bar N^{1+2/D}}
\right)}
+ 1
\right]\!,~~
\ea
where the averages are computed using the formula
\be
\langle O \rangle 
=
\left\{
\baa{lll}
\frac{2 \pi^{D/2}}{\Gamma(D/2)\, \bar N}
\int\limits_0^{\infty } \psi^* (\hat O \psi) \, r^{D-1} d r&\text{if}&D>1,\\
\frac{1}{\bar N}
\int\limits_{-\infty}^{\infty} \psi^* (\hat O \psi) d r&\text{if}&D=1,
\eaa
\right.
\ee
where $\hat O$ being a given operator.
One can see that $\xi$ is proportional to the mean-square radius of condensate,
therefore Eq. (\ref{e18}) can be viewed as an equation of the motion of a unit-mass
fictitious
particle moving in a positive $\xi$ direction
\be\label{e19}
\ddot{\xi}=
-\frac{1}{D}
\frac{d}{d \xi} U(\xi),
\ee                    
where
\be\label{e20}
U(\xi)=
D \left(
\frac{1}{2\xi^{2}}+ \ln{\xi^2}
\right)
\ee
is an effective potential,
and
\be\label{e20en}
{\cal E}
=
\frac{1}{2} 
\dot\xi^2
+
U(\xi)
\ee
is a fictitious particle's energy.
In terms of this mechanical analogy, the interpretation of Eq.~(\ref{e18}) is as follows. 
The term proportional to $\xi^{-3}$  is related to the dispersive effect 
caused by 
a spatial gradient term in Eq. (\ref{ftlagrlog}), 
while the term proportional to $\xi^{-1}$ comes from the nonlinear logarithmic term. 
The system's dynamics is thus determined by competition between these terms: at small $\xi$'s
the gradient term dominates, and at large $\xi$'s the logarithmic one does.
From the asymptotics of Eq.~(\ref{e20}), one can deduce that the logarithmic term should prevent the condensate
from unbounded spreading ($\xi \to \infty$), 
whereas the gradient and logarithmic terms together should prevent the condensate
from collapse  ($\xi \to 0$).

\begin{figure}[htbt]
\begin{center}\epsfig{figure=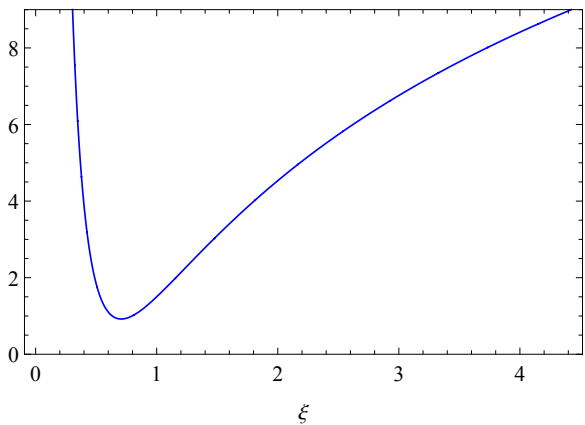,width=  1.0\columnwidth}\end{center}
\caption{
Effective potential (\ref{e20}) versus 
$\xi$, evaluated at $D=3$.  
}
\label{f:fupotlog}
\end{figure}

The potential (\ref{e20})
has a simple form, as shown in Fig. \ref{f:fupotlog}.
It has a single \textit{global} minimum and diverges at both small and large $\xi$'s, 
which means that the only allowed motion of the system is an oscillation around this minimum.
The fixed-point width of the condensate can be calculated 
from the condition
\begin{equation}\label{e23}
\left.
                       \frac{d U(\xi)}{d \xi}
\right|_{\xi=\xi_{0}}=0
,
\ee
which yields
\be\label{e26}
\xi_{0} = 1/\sqrt 2
.
\ee
Expanding Eq.~(\ref{e18}) around this fixed pint, 
we obtain the dynamical equation of the width
 \begin{equation}\label{e27}
                       \xi=\xi_{0}+ A_0\sin(\omega_{r} t+\phi_0)\,.
\ee
where  $\omega_{r}$ is a frequency of collective oscillations:
\be\label{e28}
\omega_{r}=
\sqrt{\left.\frac{1}{D}\frac{d^{2}U(\xi)}{d^{2}\xi}\right|_{\xi=\xi_{0}}}
=
\frac{1}{\xi_{0}^2}
\sqrt{3 -2 \xi_{0}^2}
,
\ee
and $A_0$ and $\phi_0$ are real-valued integration constants.
The frequency $\omega_{r}$ can be used to analyze the stability of the condensate:
the solution (\ref{e26}) is stable only if frequencies of
collective modes are real-valued.
In our case, using Eq.~(\ref{e26}), one obtains
\be\label{e29}
\omega_{r}=
2 \sqrt 2
,
\ee
which indicates that our solution is indeed stable, without any critical points.

\sscn{Vakhitov-Kolokolov stability}{s:ssvk}

Another criterion for stability is the Vakhitov-Kolokolov one \cite{vk73},
which in our case reads \cite{r24}:
\be
\frac{\partial  N}{\partial \mu}
< 0
,
\ee 
assuming our notation conventions for this section.

In order to determine the chemical potential, let us find the ground state of our system
(an importance of studying the ground state's stability is discussed in Sec. \ref{s:nst} above).
One can derive that an exact ground-state solution of Eq. (\ref{e7})
is given by a Gaussian:
\be
\phi_0 (r)
=
\pm
\sqrt{\bar N}
\left(
2 / \pi
\right)^{D/4}
\exp{(- r^2)}
,
\ee
while Eq. (\ref{e7}) reduces to an algebraic equation for the eigenvalue $\mu$.
Solving it, we obtain
\be
\mu_0 =
\ln{\left(\bar N_c^2 / \bar N^2 \right)}
=
\ln{\left(N_c^2 / N^2 \right)}
,
\ee
where we denoted the critical value
\be
\bar N_c
=
N_c
(a/\ell)^D
=
(\pi \text{e}^2/2)^{D/2} 
,
\ee
which corresponds to the number of particles 
at which the chemical potential changes its sign.
Using these formulae, we obtain
\be
\frac{\partial  N}{\partial \mu_0}
=
- \frac{1}{2}  N
<
0
,
\ee
which means that the trapless logarithmic condensate is also VK-stable.
Moreover, its formation is energetically favorable for $N > N_c$.

To summarize, both approaches have shown us that 
the trapless $D$-dimensional logarithmic BEC is stable, even in absence of any trapping potentials,
which makes it unique among all other known condensates which require external potentials for 
stability (cf. Sec. \ref{s:poly}).
These results confirm an earlier idea \cite{r21} that
the logarithmic condensate behaves more like a liquid than a gas - for instance, 
in the absence of any forces including gravity, it should form a Gaussian droplet
which stability was demonstrated in Ref. \cite{r20}
and
recently confirmed by means of an orbital stability approach \cite{ard16}.
The stability of such a droplet is ensured not by 
surface tension but by quantum nonlinear effects in its bulk.

\scn{Trapless BEC with 
few-body interactions}{s:poly}

For the sake of comparison with a logarithmic case, let us study 
trapless BEC with few-body interactions
that evolves in 
the $D$-dimensional Euclidean space which is
free of any external potentials
including gravity.
We begin by considering an
isotropic BEC with both two- and three-body interactions.  
The formalism of Refs. \cite{r24,r22,r28}, 
where the $D$-dimensional condensate with two- and three-body interactions was considered
in a harmonic trap,
will be used in this section.
However, 
those results alone cannot be directly applied to a completely trapless case -- because
in a zero trap frequency limit, some parameters used become singular.
Thus, for a trapless BEC, one should start derivations anew,
similar to what was done for a 3D case \cite{ackm03,adh04,svpm10}.

The wave equation for the condensate  
with two- and three-body interactions at zero temperature 
takes a form
of the cubic-quintic Schr\"odinger equation:
\be\label{ebody23}
i \hbar \frac{\partial}{\partial t}\Psi
=
\left[
-\frac{\hbar^{2}}{2 m} \lapl
+\frac{\lambda_{2}}{2}
 |\Psi|^{2}
+\frac{\lambda_{3}}{2}
 |\Psi|^{4}
\right]\Psi
,
\ee
where the condensate wavefunction is normalized as in Eq. (\ref{norm}),
$\lambda_{2}$ and $\lambda_{3}$ are real coupling constants (we do not consider dissipative effects here).
The corresponding Lagrangian density is
given by a formula analogous to Eq. (\ref{ftlagrlog}) 
where the field-theoretical potential density 
is defined as
\be\lb{e:ftpotcq}
V (n) 
=
\frac{1}{2}
\sum\limits_{k=2}^3
\frac{\lambda_{k}}{k} n^k
= 
\frac{\lambda_{2}}{4} n^{2} +\frac{\lambda_{3}}{3!}  n^{3}
,
\ee
for $n = |\Psi|^2$.

Furthermore, because (effectively) one- and two-dimensional condensates are impossible to contain 
without some kind of trapping potential or geometric constraint 
(which is \textit{de facto} a trapping potential too), we can
restrict ourselves to the case 
\be
D > 2
,
\ee
while the lower-dimensional cases can be considered by analogy.
Then the system (\ref{ebody23})
has three natural scales of length,
three of time, and two of mass:
\ba
\text{L} 
&=& 
\left\{ 
\left(\frac{m |\lambda_2|}{\hbar^2}\right)^{\frac{1}{D-2}},  
\left(\frac{m |\lambda_3|}{\hbar^2}\right)^{\frac{1}{2(D-1)}}, 
\left(\frac{|\lambda_3|}{|\lambda_2|}\right)^{\frac{1}{D}}
\right\}
,\nn\\
\text{T} 
&=& 
\left\{ 
\left(\frac{m^D \lambda_2^2}{\hbar^{D+2}}\right)^{\frac{1}{D-2}}, 
\left(\frac{m^D |\lambda_3|}{\hbar^{D+1}}\right)^{\frac{1}{D-1}}, 
\hbar \frac{|\lambda_3|}{\lambda_2^2}
\right\}
,~~\\
\text{M} 
&=& 
\left\{m, 
\hbar^2
\left( \frac{|\lambda_3|^{D-2}}{\lambda_2^{2(D-1)}} \right)^{\frac{1}{D}}
\right\}
,\nn
\ea
which can be used to obtain dimensionless quantities.
Assuming $\lambda_2 \not= 0$
and introducing the notations
\ba
&&
\ell_2 = \left(\frac{m |\lambda_2|}{\hbar^2}\right)^{\frac{1}{D-2}}
, \
\tau_2 = \left(\frac{m^D \lambda_2^2}{\hbar^{D+2}}\right)^{\frac{1}{D-2}}
,\\&&
\ell_3 = \left(\frac{|\lambda_3|}{|\lambda_2|}\right)^{\frac{1}{D}}
, \
\tau_3 = \hbar \frac{|\lambda_3|}{\lambda_2^2}
,
\ea
and
\be
\vec r \,'= \vec r/\ell_2,\
t' = t/\tau_2
,\ 
\psi = \ell_3^{D/2} \Psi
,
\ee
we can write Eq. (\ref{ebody23}) in a dimensionless form:
\be\label{e55}
i \partial_{t'} \psi
=
-
\frac{1}{2} \lapl^\prime \psi
+
\frac{\becc}{2}
\left(
s_2 |\psi|^{2}
+
s_3 |\psi|^{4}
\right)
\psi
,
\ee
where $\becc = \tau_2/\tau_3$, and $s_k = \text{sgn}(\lambda_k) = \lambda_k/|\lambda_k|$.
Besides, the normalization condition reads
\be\lb{normcq}
\left\langle \psi | \psi\right\rangle'
\equiv
\int |\psi|^2 d^D \vec{r}\,'  
=
N (\ell_3/\ell_2)^{D} \equiv \tilde N
,
\ee
where $\tilde N$ is the reduced number of particles of the CQSE condensate.
Since $\becc$ is non-negative by construction, signs $s_k$ define the type of 
corresponding $k$-body interaction: repulsive (plus) or attractive (minus). 

In this section, we do not consider the linear or orbital criteria of stability:
even though nontrivial solutions of Eq. (\ref{e55}) do exist, 
they correspond to excited states of the CQSE system, therefore they will be unstable
against spontaneous quantum transitions, as discussed in Sec. \ref{s:nst}.
Those transitions will eventually bring the system to its ground state -- which is a 
trivial one, $\psi_0 = 0$.
In what follows, we study only the variational stability of the system,
for which one does not need to know any solution of Eq. (\ref{e55}).

From now on, we omit primes assuming that in this section, a length is measured in
units $\ell_2$, time - in units $\tau_2$, energy - in units $\hbar/\tau_2$, and so on.
Using the formalism of Sec. \ref{s:ssva}, 
including the notations for the trial function (\ref{e8}), 
we can write 
the averaged Lagrangian in the form
\ba
L
&=&
\pi^{D/2}A^2\xi^{D}
\Biggl[
\frac{D}{4\xi^2}
+\frac{D}{2} \xi^2\left(\dot{\beta}+2\beta^2\right)
+\dot{\alpha}
\nn\\&&
+
\frac{\lambda}{2}
\left(
\frac{s_2  A^2}{2^{1+D/2}}
+ 
\frac{s_3 A^4}{3^{1+D/2}}
\right)
\Biggr]
,
\label{e11cq}
\ea
where 
the dot denotes a time derivative.
By analyzing the corresponding Euler-Lagrange equations,
we obtain
\ba
\dot{\beta}
&=&
\frac{\lambda \tilde N}{4\xi^{2}}
\left(
\frac{s_2}{(2\pi)^{D/2} \xi^{D}}
+ 
\frac{4 s_3 \tilde N}{\pi^{D} 3^{1+D/2} \xi^{2D}}
\right)
\nn\\&&
+ \frac{1}{2\xi^4}
- 2 \beta^2
,\label{e15cq}\\
\dot{\xi}
&=&
2\xi\beta
, \label{e16cq}
\ea
and
\begin{equation}
\tilde N =
\pi^{D/2} A^2 \xi^D
= \text{const}
,
\label{e9cq}
\ee
assuming the definition (\ref{normcq}).
From these equations one can easily derive
the width equation and fictitious particle's effective potential:
\ba
&&
\ddot{\xi}=\frac{1}{\xi^{3}}
+\frac{s_2 P}{\xi^{D+1}}+\frac{s_3 Q}{\xi^{2 D +1}}
,\label{e61}\\&&
U(\xi)=
\frac{D}{2 \xi^{2}}+\frac{s_2 P}{\xi^{D}}+\frac{s_3 Q}{2\xi^{2 D}}
,\label{e62}
\ea
where we have introduced the dimensionless magnitudes of two- and three-body interactions:
\[
P= \frac{\becc \tilde N}{2(2\pi)^{D/2}},
		\
Q= \frac{2 \becc\tilde N^{2}}{3^{1+D/2}\pi^{D}}
= \frac{2^{1+D/2} \tilde N P }{3 (3 \pi)^{D/2}}
= \frac{2^{1+D} P^2 }{3^{1+D/2} \becc}
		,
\]
and the fictitious particle's energy ${\cal E}$ is given
by a general formula (\ref{e20en}).

Since we consider a case $D>2$,
the potential $U(\xi)$ has the following asymptotics:
\ba
\xi\to 0&:& \ \ U (\xi) 
\to
\left\{
\baa{lll}
s_3 Q/(2\xi^{2 D}) & \text{if} & Q\not= 0,\\
s_2 P/\xi^{D} & \text{if} & Q = 0,
\eaa 
\right.
\lb{asy0}\\
\xi\to +\infty &:& \ \ U (\xi) 
\to 0
,
\lb{asyi}
\ea
which indicates that it always diverges in the $\xi$-origin but vanishes at spatial infinity, cf. Fig. \ref{f:tt}.
This means that,
depending on values $P$, $Q$ and $s_k$, either $U(\xi)$ has no fixed points 
at $0 < \xi < \infty$, 
or those points are extrema, hence a finite value of energy ${\cal E}$ 
always exists at which the fictitious particle eventually escapes
to $\xi$-infinity or hits the origin $\xi = 0$.

\begin{figure}
\centering
\subfloat[$s_2 = -1,\, s_3=-1$]{
  \includegraphics[width=\sct\columnwidth]{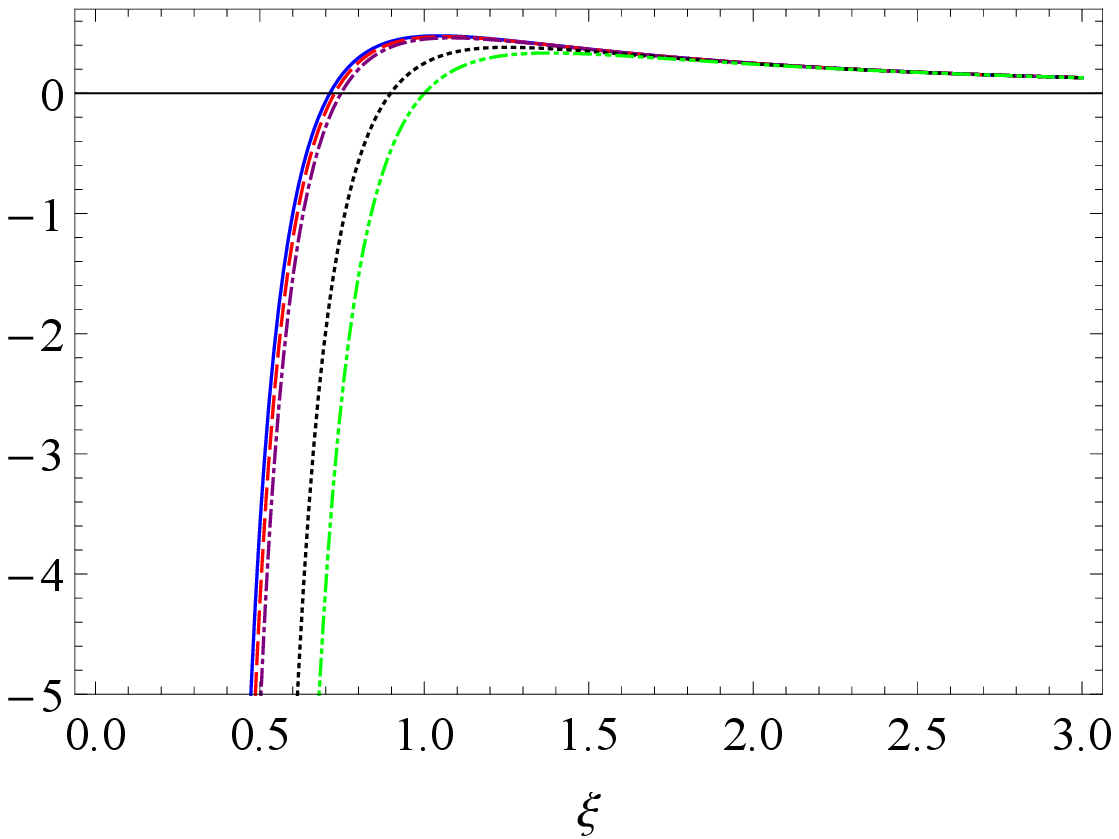}
}
\subfloat[$s_2 = -1,\, s_3=1$]{
  \includegraphics[width=\sct\columnwidth]{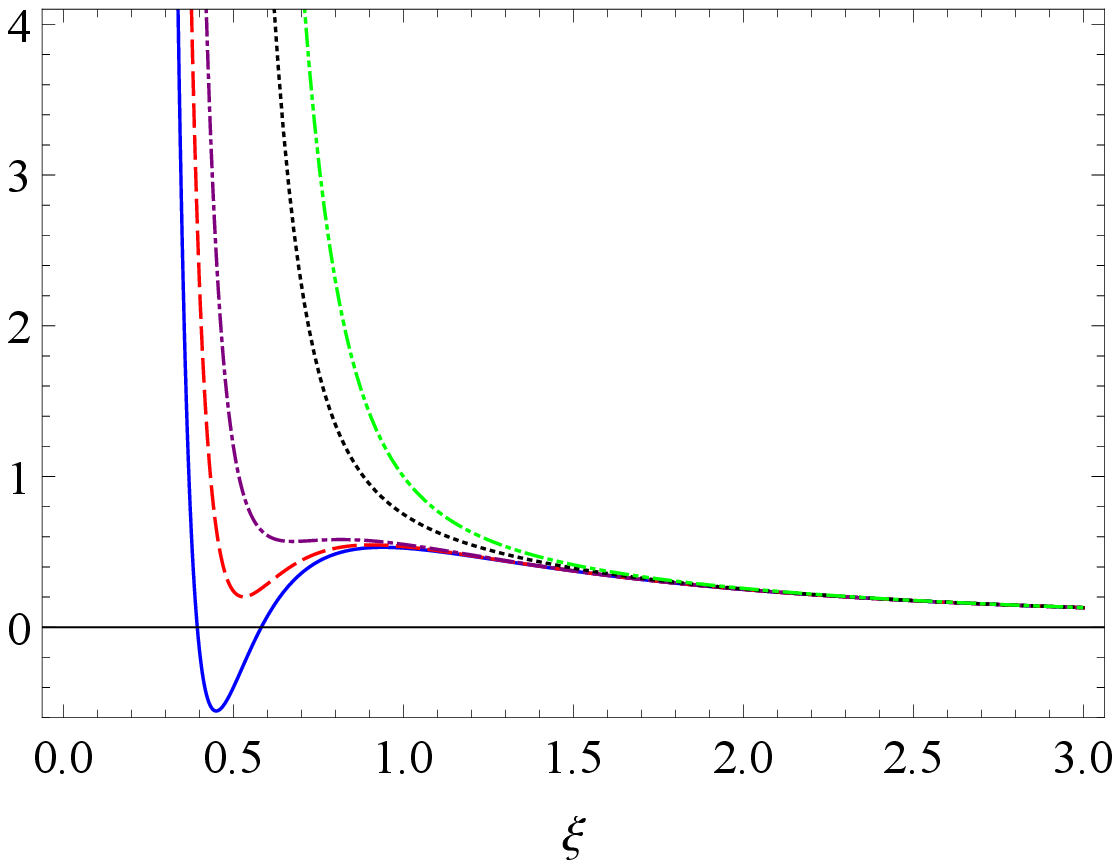}
}
\hspace{0mm}
\subfloat[$s_2 = 1,\, s_3=-1$]{
  \includegraphics[width=\sct\columnwidth]{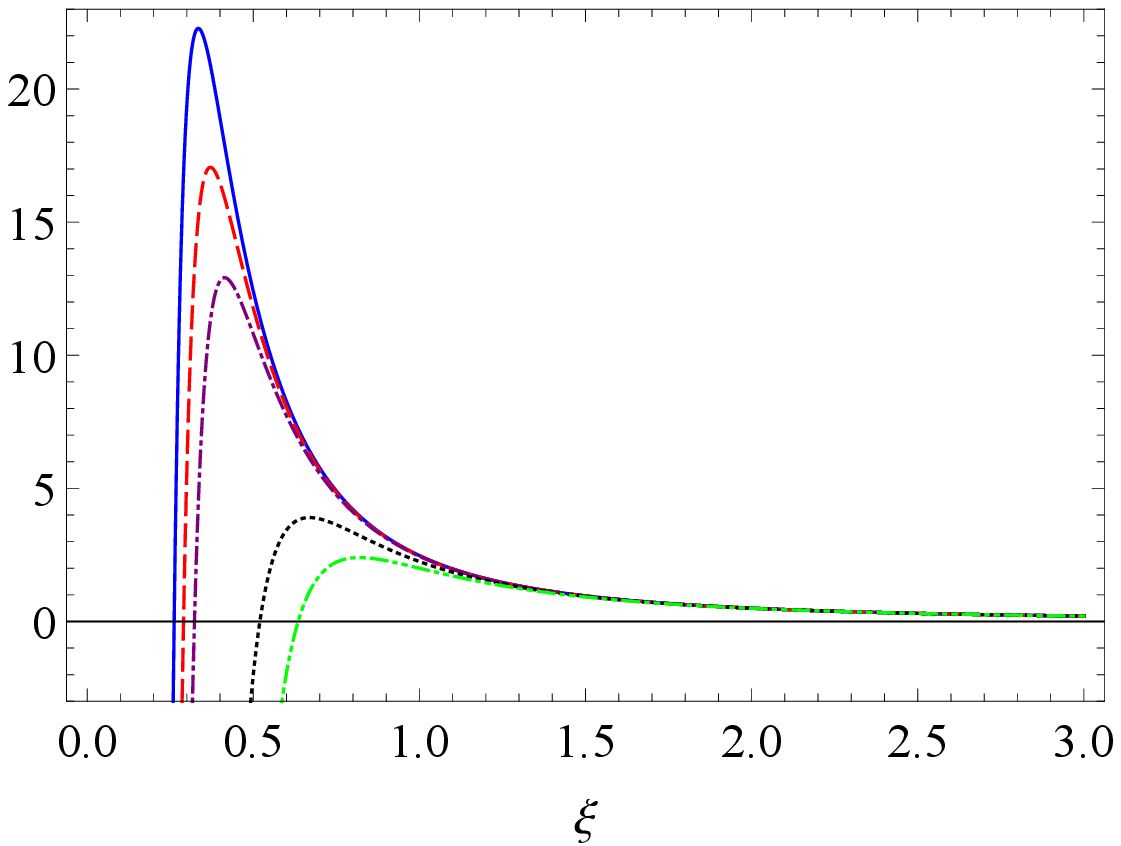}
}
\subfloat[$s_2 = 1,\, s_3=1$]{
  \includegraphics[width=\sct\columnwidth]{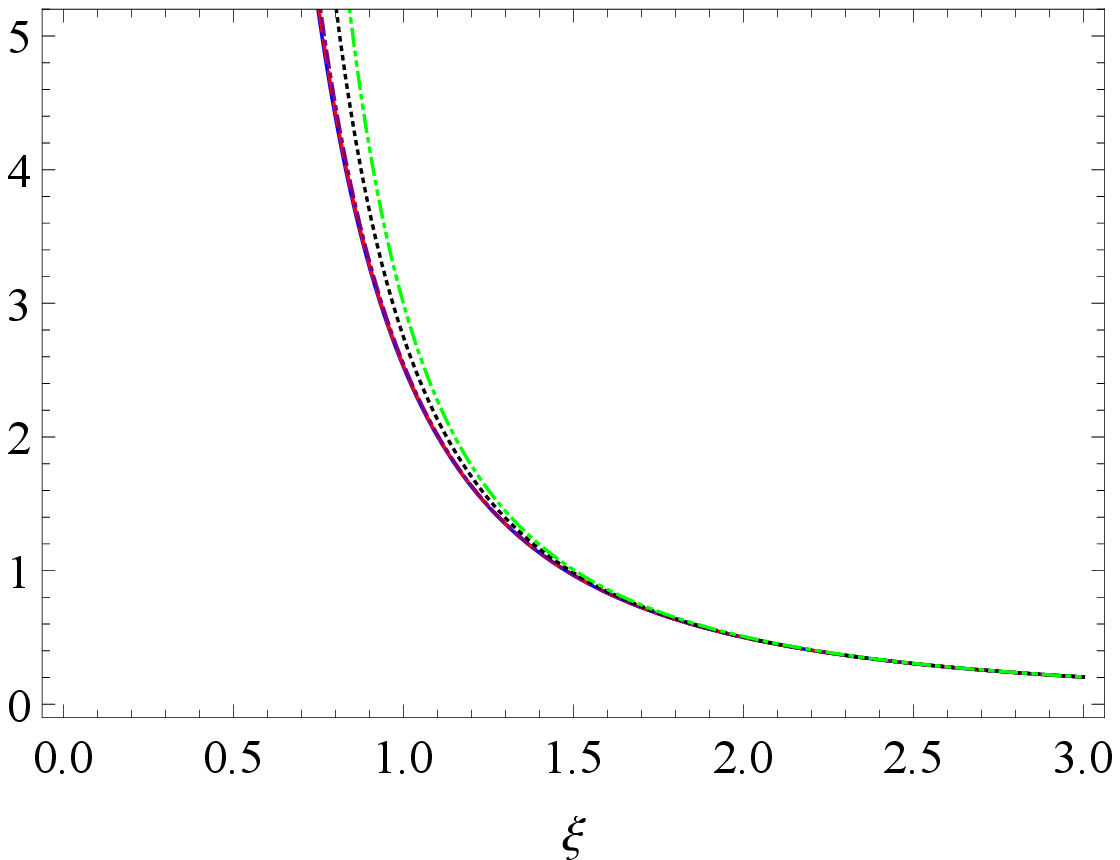}
}
\caption{
Effective potential (\ref{e62}) versus 
$\xi$, evaluated at $D=3$, $P=1$, and 
different values of $Q$:
$0.05$ (solid curves),
$0.07$ (dashed curves),
$0.1$ (dash-dotted curves),
$0.5$ (dotted curves)
and
$1$ (dash-double-dotted curves).
The upper (lower) row of panels corresponds to an attractive (repulsive) two-body interaction,
while the left (right) column corresponds to an attractive (repulsive) three-body interaction.
}
\label{f:tt}
\end{figure}

In physical terms, this means that the trapless CQSE condensate  
can be, in the best case scenario, metastable:
even if it is stable against the collapse ($\xi \to 0$) it is 
unstable against delocalization (``spreading'').
The latter can be \textit{dynamic} if the effective potential (\ref{e62})
has no local minima at $0 < \xi < \infty$, cf. Figs. \ref{f:tt}a, \ref{f:tt}c or \ref{f:tt}d,
or \textit{spontaneous} if the potential 
has at least one minimum at $0 < \xi < \infty$, cf. the solid or dashed curve
in Fig. \ref{f:tt}b.  

Within the frameworks of the variational approach, spontaneous delocalization
occurs due to fluctuations of a condensate's width $\xi (t)$, which are
inevitable in the quantum realm.
It 
can be effectively described
as macroscopic tunneling of a fictitious
particle towards $\xi$ infinity through a finite
potential barrier,
$
\Delta U =
 U (\xi_\text{max}) - U (\xi_\text{min})
,
$
where 
$
0 <
\{
\xi_\text{max}, \, \xi_\text{min}
\}
< \infty 
$ are the local extrema's points for the potential $U (\xi)$ \cite{sto97}.
Note that this process should not be confused with the quantum tunneling
of trapped Bose-Einstein condensates through a trap potential located in the 
configuration space $\vec r$: here we work in terms of a collective degree of freedom $\xi (t)$
and a number of particles $N$ is conserved.
It should also be noted that energy ${\cal E}$ 
of a fictitious particle in $\xi$ space is not the same 
as energy of a condensate in the configuration space.

The transmission coefficient at a given energy ${\cal E}$ 
can be easily computed in a semiclassical approximation as
\be
T (\en) = 
\exp{ \left(-2 
K_{2 3}
\right)
}
,
\ee
where
\be
K_{j k}
\equiv
\int\limits_{\xi_j}^{\xi_k}\!
\sqrt{2(U (\xi) - \en)}
\,
d \xi
,
\ee
and $\xi_2$ and $\xi_3$ are classical turning points in a region under the barrier,
$\xi_2 < \xi_\text{max} < \xi_3$.
The lifetime $\tau_\xi$
of the condensate which undergoes spontaneous delocalization
can be easily computed in a semiclassical approximation.
Assuming that the tunneling 
probability through the barrier 
$
\Delta U
$
is small and therefore 
\be
K_{2 3} \gg 1
,
\ee
we obtain
\be\lb{elifet}
\tau_\xi
\approx
\frac{2}{T(E)} 
\left(
\frac{\partial J_{1 2}}{\partial E}
\right)_{E = E_n}
,
\ee
where the
value $E_n$ is a real-valued solution of the eigenvalue equation
\be
J_{1 2} - \pi (n + 1/2) = 0
,
\ee
$n$ being an integer,
and
\be
J_{j k}
\equiv
\int\limits_{\xi_j}^{\xi_k}\!
\sqrt{2(\en - U (\xi))}
\,
d \xi
,
\ee
and $\xi_1$ and $\xi_2$ are classical turning points in the adjacent well on the left-hand side from the barrier,
$\xi_1 < \xi_\text{min} < \xi_2< \xi_\text{max} < \xi_3$.
Similarly, one can derive the characteristics of macroscopic tunneling 
towards the $\xi$ origin, if it is allowed by the effective potential's form.

As a result, 
the trapless CQSE condensate always has a finite lifetime (except in those 
cases when a potential $U (\xi)$ has a global minimum at a finite positive $\xi$,
with at least one negative energy ${\cal E}$ level, 
cf. a solid curve in Fig. \ref{f:tt}b,
and energy fluctuations are somehow suppressed):
it
tends to either occupy all the available volume (hence get depleted)
or collapse to a state with a delta-singular density profile.
In other words, 
it is unstable against either unrestricted expansion (hence dilution) or collapse,
depending on values $P$, $Q$ and attraction/repulsion indicators $s_k$.
Such a metastability can be easily seen in reality: 
models like (\ref{ebody23}) are known to be applicable for gaseous condensates, 
therefore, some kind of trapping potential or geometrical constraint
would be necessary for their ``eternal'' stability,
otherwise the system quickly depletes with time.
As for the collapse process, in practice it
stops at a length scale for which condensed atoms can no longer be regarded as point-like Bose particles,
or the few-body approximation becomes no longer applicable.

Now let us consider a trapless BEC with arbitrary few-body interactions.
Most of above-mentioned features remain valid --
since in a minimally-coupled $U(1)$-symmetric case,
such a condensate would be described by some kind of polynomially nonlinear Schr\"odinger equation,
\be\label{ebodyfew}
i \hbar \frac{\partial}{\partial t}\Psi
=
\left[
-\frac{\hbar^{2}}{2 m} \lapl
+
\frac{1}{2}
\sum\limits_{k=2}^\nbod
\lambda_{k} |\Psi|^{2(k-1)}
\right]\Psi
,
\ee
the few-body analogue of Eqs. (\ref{e61}) and (\ref{e62}) would be, respectively:
\ba
&&
\ddot{\xi}=
\frac{1}{\xi^{3}}
+
\frac{1}{\xi}
\sum\limits_{k=2}^\nbod
\frac{I_k}
{\xi^{
(k-1)
D}
}
,\lb{e61f}
\\&&
U(\xi)
=
\frac{D}{2\xi^{2}}
+
\sum\limits_{k=2}^\nbod
\frac{I_k}
{(k-1)\xi^{
(k-1)
D}
}
,
\lb{e61p}
\ea
where the coefficient $I_k = I_k (\lambda_k)$ is a function of 
a $k$-body interaction strength parameter, $2 \leqslant \nbod < \infty$
is a maximum amount of particles that can interact simultaneously,
and the fictitious particle's energy ${\cal E}$ is given
by a general formula (\ref{e20en}).
The asymptotic properties of Eq. (\ref{e61p}), 
\ba
\xi\to 0&:& \ \ U (\xi) 
\to
\frac{I_\nbod}{(\nbod-1) \xi^{(\nbod-1) D}}
,
\lb{asy0f}\\
\xi\to +\infty &:& \ \ U (\xi) 
\to 0
,
\lb{asyif}
\ea
are
qualitatively similar to 
Eqs. (\ref{asy0}) and (\ref{asyi}).
As shown above, this implies, at least, the suppression of stability against 
unbounded expansion caused by spontaneous delocalization, due to the presence
of the width's fluctuations.
This means that the trapless few-body condensates described 
by ``polynomial'' models are at best metastable,
with a finite lifetime determined by Eq. (\ref{elifet}),
except in some cases when the effective potential (\ref{e61p}) has a global minimum at $0 < \xi < \infty$,
with at least one negative level of energy ${\cal E}$, which can stabilize the system 
(in absence of energy fluctuations).

\scn{Conclusion}{s:con}

In the present work,
the stability of a trapless condensate described by the logarithmic Schr\"odinger
equation was studied 
and compared with a case of a trapless BEC with few-body interactions,
described by wave equations with polynomial nonlinearity,
such as GPE or CQSE.

By arguing that one can always expand transcendental functions, such as logarithm,
into Taylor series, one might expect that the properties of the logarithmic model would be
similar to the few-body models, at least qualitatively, and that the few-body models' properties 
could reproduce all the features of the logarithmic model by considering sufficiently
many terms in the series expansion.
However, we showed that these assumptions are incorrect, in general: 
by restricting oneself to any finite number of terms in series, one drastically changes the main properties of a corresponding condensate model.
In other words, a nonperturbative treatment is essential when dealing with 
the ``transcendental'' condensates in general and the logarithmic ones in particular.

The Gaussian variational approach and Vakhitov-Kolokolov criterion
were used to determine the dynamics and stability of the logarithmic condensate, in $D$ dimensions. 
Using natural symmetry assumptions,
we derived the collective oscillations frequency and the mean-square radius of the condensate. 
Further, it was demonstrated
that the trapless logarithmic 
condensate is always stable -- essentially, because logarithmic
nonlinearity prevents it from both the collapse and unbounded expansion (hence dilution).

One notices that, according to Eqs. (\ref{e19})-(\ref{e26}),
trapless logarithmic Bose-Einstein condensate is attractive if its width 
is above a certain length scale, 
approximately
$\xi_0\ell$, 
and
repulsive if it is below.
Therefore, it can be used for modeling bosenova-type phenomena when the Bose-Einstein condensate 
shrinks to a size smaller than the minimum resolution limit of a detector, and then rapidly expands.

Finally, stability studies of trapless condensates with few-body interactions,
described by polynomially nonlinear wave equations,  
demonstrated that such condensates are unstable against unbounded expansion or collapse,
unless one applies an external potential or geometric constraint to them.
The crucial indicator here is the shape of their effective potential $U (\xi)$
which governs the dynamics of collective oscillations in terms of the width $\xi (t)$,
a collective degree of freedom of a condensate.
It is generally shown that:
(i) if this potential has neither confining shape nor local minima then the condensate is dynamically unstable
against delocalization,
(ii) if this potential does not have a confining shape (\textit{e.g.}, it vanishes at infinity)
but has
at least one local minima then the condensate is metastable, \textit{i.e.}, unstable
against the spontaneous delocalization,
and thus has a finite lifetime
(except in some special
cases
discussed below).

By comparing these features to the logarithmic case (for which the effective potential does have an
absolute minimum and confining shape, cf. Fig. \ref{f:fupotlog}),
one can deduce 
that ``transcendental'' condensates, such as the logarithmic one, can be used for 
describing stable quantum liquids (which was indeed shown in the work \cite{r21}), 
while ``polynomial'' ones are \textit{a priori} more suitable for describing 
low-density quantum matter,
such as diluted cold gases
(although, even there their applicability might have limits, as indicated by experiments \cite{khm17}).
Besides,
a special class 
of trapless ``polynomial'' condensates exists,
for which the effective potential $U (\xi)$ vanishes at infinity,
has a global minimum at a finite positive $\xi$, and allows 
at least one negative level of effective energy ${\cal E}$.
In this case, the condensate would be stable in absence of energy fluctuations
(regardless of the presence of width fluctuations),
but even a small increase of energy can 
excite the system into a metastable state with a nonzero probability of delocalization.
Such models can be used for 
describing those condensates, which stay localized, similarly to liquids, in absence of energy fluctuations,
but expand like gases otherwise.

\begin{acknowledgments}
Fruitful discussions with participants of the International Workshop ``Symmetry and Integrability of 
Equations of Mathematical Physics'' (17-20 December, 2016, Institute of Mathematics of National Academy of Sciences of Ukraine, Kyiv), 
where parts of this work were presented, 
are acknowledged.
This work is based on the research supported by the National Research Foundation of South Africa 
under Grants Nos.
95965 
and 98892.
Proofreading of the
manuscript by P. Stannard is greatly appreciated.

\end{acknowledgments}


\end{document}